\documentclass[aps,prl,preprint,superscriptaddress]{revtex4-1}

\usepackage{graphicx}
\usepackage[english]{babel}
\usepackage{amsmath,amssymb,amsfonts,latexsym,amsbsy}
\usepackage{gensymb}
\usepackage[colorlinks=true,linkcolor=blue,citecolor=blue,urlcolor=blue]{hyperref}
\usepackage{xcolor}
\usepackage{lineno}
\usepackage{siunitx}
\usepackage{url}
\usepackage{tikz}

\definecolor{lime}{HTML}{A6CE39}
\DeclareRobustCommand{\orcidicon}{
	\begin{tikzpicture}
	\draw[lime, fill=lime] (0,0) 
	circle [radius=0.16] 
	node[white] {{\fontfamily{qag}\selectfont \tiny ID}};
	\draw[white, fill=white] (-0.0625,0.095) 
	circle [radius=0.007];
	\end{tikzpicture}
	\hspace{-2mm}
}

\foreach \x in {A, ..., Z}{\expandafter\xdef\csname orcid\x\endcsname{\noexpand\href{https://orcid.org/\csname orcidauthor\x\endcsname}
			{\noexpand\orcidicon}}
}

\addto\captionsenglish{}
\addto\captionsenglish{}

\begin{document}


\title{Origin of Mechanical and Dielectric Losses from Two-Level Systems in Amorphous Silicon}


\author{M. Molina-Ruiz\orcidA{}}
\email[Corresponding author: ]{manelmolinaruiz@gmail.com}
\affiliation{Department of Physics, University of California Berkeley, Berkeley, CA 94720}

\author{Y. J. Rosen\orcidB{}}
\affiliation{Lawrence Livermore National Laboratory, Livermore, CA 94550}

\author{H. C. Jacks}
\altaffiliation[Present address: ]{California Polytechnic University, Physics Department, San Luis Obispo, CA, 93407}
\affiliation{Department of Physics, University of California Berkeley, Berkeley, CA 94720}

\author{M. R. Abernathy\orcidD{}}
\altaffiliation[Present address: ]{The Johns Hopkins University Applied Physics Laboratory, Laurel, MD, 20723}
\affiliation{Naval Research Laboratory, Code 7130, Washington, D.C. 20375}

\author{T. H. Metcalf}
\affiliation{Naval Research Laboratory, Code 7130, Washington, D.C. 20375}

\author{X. Liu\orcidG{}}
\affiliation{Naval Research Laboratory, Code 7130, Washington, D.C. 20375}

\author{J. L Dubois}
\affiliation{Lawrence Livermore National Laboratory, Livermore, CA 94550}

\author{F. Hellman\orcidI{}}
\affiliation{Department of Physics, University of California Berkeley, Berkeley, CA 94720}
\affiliation{Lawrence Berkeley National Laboratory, Berkeley, CA 94720}


\date{\today}

\begin{abstract}
Amorphous silicon contains tunneling two-level systems, which are the dominant energy loss mechanisms for amorphous solids at low temperatures. These two-level systems affect both mechanical and electromagnetic oscillators and are believed to produce thermal and electromagnetic noise and energy loss. However, it is unclear whether the two-level systems that dominate mechanical and dielectric losses are the same; the former relies on the coupling between phonons and two-level systems, with an elastic field coupling constant, $\gamma$, while the latter depends on a two-level systems dipole moment, $p_0$, which couples to the electromagnetic field. Mechanical and dielectric loss measurements as well as structural characterization were performed on amorphous silicon thin films grown by electron beam deposition with a range of growth parameters. Samples grown at 425 \degree C show a large reduction of mechanical loss (34 times) and a far smaller reduction of dielectric loss (2.3 times) compared to those grown at room temperature. Additionally, mechanical loss shows lower loss per unit volume for thicker films, while dielectric loss shows lower loss per unit volume for thinner films.  Analysis of these results indicate that mechanical loss correlates with atomic density, while dielectric loss correlates with dangling bond density, suggesting a different origin for these two energy dissipation processes in amorphous silicon.
\end{abstract}


\maketitle

\section{Introduction}
Amorphous insulators exhibit anomalous elastic and dielectric responses to external fields at low temperatures due to quantum tunneling between nearly-degenerate states~\cite{Zeller1971,Hunklinger1972,Stephens1973,Schickfus1975}. These tunneling states, which are generally approximated by two-level systems (TLSs), are reasonably well described by the standard tunneling model (STM)~\cite{Anderson1972,Phillips1972}. However, there are significant gaps in the theory which could be the key to ameliorating the effects of TLSs on low temperature technologies, such as quantum decoherence and noise in superconducting quantum hardware~\cite{Martinis2005,Gao2008,Muller2019}. The STM describes the interaction between applied external elastic fields and TLSs, which interact by means of the deformation potential or coupling constant, $\gamma$, and cause energy dissipation or mechanical loss. The STM also describes the interaction with electric fields via the electric dipole moment, $p_0$, which occurs due to a charge reconfiguration between the different particle rearrangements in the two states, and as a consequence part of the energy is dissipated, the dielectric loss. It can be argued that due to the atomic rearrangement needed to provide a dipole moment, the TLSs that couple to electric fields should also couple to elastic fields. The converse, that systems responsible for mechanical response must be associated with an electric response, is not necessarily true. In fact, Arnold et al.~\cite{Arnold1976} found that they could reduce the dielectric loss in borosilicate glass, but they only affected an estimated half of the TLSs that were elastically coupled. Dielectric loss is generally dominated by TLSs with large electric-field coupling, while the mechanical loss is determined by TLSs that strongly couple to elastic fields. It is not clear whether these two species of TLSs are the same, whether there is a relationship between the two types of loss, and whether there are predictors for the expected dissipation rates of a material.

In amorphous solids, energy dissipation is generally used as a measure of the density of TLSs, $\bar{P}$. However, to the best of our knowledge, there has not been a systematic study of the relationship between mechanical and dielectric loss. While both elastic and electric coupling mechanisms are generally well described by the STM~\cite{Golding1976,Hunklinger1976,Schickfus1977,Golding1979}, it is not clear whether TLSs responsible for the different phenomena are the same, or whether they are even correlated.

In our previous work, we showed that TLS density derived from low temperature mechanical loss and from excess low temperature specific heat is greatly affected by growth temperature and thickness in amorphous silicon ($a$-Si) films, and that this dependence was explained by a strong dependence on atomic density; by contrast, elastic properties such as sound velocity and shear modulus depend only on growth temperature~\cite{Queen2013,Liu2014,Queen2015}. These two facts taken together led to the suggestion that elastic waves are carried by a continuous high density medium, while TLSs originate in lower density regions. Recently, we have shown that reduced atomic density in $a$-Si films is associated with low density regions, i.e., nanovoids~\cite{Jacks2018,Jacks2020}. Silicon is a fourfold coordinated atom that preferentially bonds to four adjacent silicon atoms and, in its crystalline form, tetrahedrally bonds over long range. Amorphous silicon lacks periodicity but preserves local tetrahedral coordination, and atoms form a continuous random network where not all atoms are tetrahedrally bonded. Some of those atoms exhibit unpaired electrons, or dangling bonds, a type of structural defect that is also a common electronic defect in $a$-Si~\cite{Stutzmann1989}. In this work, we study mechanical and dielectric loss on $a$-Si films grown at different temperatures and thicknesses and show that the mechanisms responsible for the two types of dissipation processes are independent. Specifically, we show that for $a$-Si, mechanical loss correlates with atomic density, while dielectric loss correlates with dangling bond density.

\section{Experimental Methods}
Amorphous silicon samples were grown by electron beam (e-beam) evaporation at a base pressure of $10^{-9}$ Torr. Samples of different thicknesses were grown at 0.5 \AA/s and three different growth temperatures $\text{T}_\text{S} =$ 50, 225, and 425 \degree C. Extensive structural characterization (high-resolution transmission electron microscopy, electron diffraction, fluctuation electron microscopy, and Raman spectroscopy) has been done on these $a$-Si films, which shows them to be completely amorphous, with systematic dependencies of atomic density on growth temperature, rate, and thickness~\cite{Queen2015,Jacks2018,Jacks2020}. Atomic density was determined by Rutherford backscattering spectrometry (RBS) in combination with thickness measurements from profilometry, and converted to mass density by multiplying by the Si atomic mass. Hydrogen within the samples is below detection limit ($< 0.1\%$) as measured by Hydrogen Forward Scattering. Samples were grown on undoped, (100) silicon with resistivity greater than 10 k$\Omega\,$cm. Substrates were chemically cleaned and baked under vacuum at 150 \degree C prior to deposition. The thin native oxide layer was left on all substrates to prevent epitaxial Si growth due to direct contact to the c-Si substrate~\cite{Sakai1963,Sorianello2011}.

Mechanical loss measurements were performed from 0.3 to 100 K using microfabricated double-paddle oscillators (DPOs)~\cite{White1995,Liu1998} with either 60 or 300 nm thick films of $a$-Si deposited at various temperatures. This technique measures the energy dissipation of transverse modes of the oscillator neck by measuring the second antisymmetric resonance mode (AS2) at approximately 5500 Hz. For simplicity in the equations below we omit the polarization subscript since all modes in the DPO measurements are transverse. The film internal friction or mechanical loss, $Q_m^{-1}$, is obtained by measuring the ring-down time (equivalent to the change in resonance frequency width) of the oscillator’s response at resonance before and after the film is deposited:
\begin{equation}
    Q_m^{-1} = \frac{G_{sub}t_{sub}}{3G_{film}t_{film}} ( Q_{total}^{-1} - Q_{bare}^{-1} )
    \label{equation1}
\end{equation}
where $G$ and $t$ refer to the shear modulus and thicknesses of the substrate or DPO (sub) and sample (film). We use $G_{sub} = 62$ GPa. The film shear modulus, $G_{film}$, is determined from the resonant frequency shift of the antisymmetric mode before and after the film has been deposited. We note that $Q_{total}^{-1}$ and $Q_{bare}^{-1}$ are extensive properties, whereas $Q_m^{-1}$ is an intensive property.

\begin{figure}
    \centering
    \includegraphics[scale=0.7]{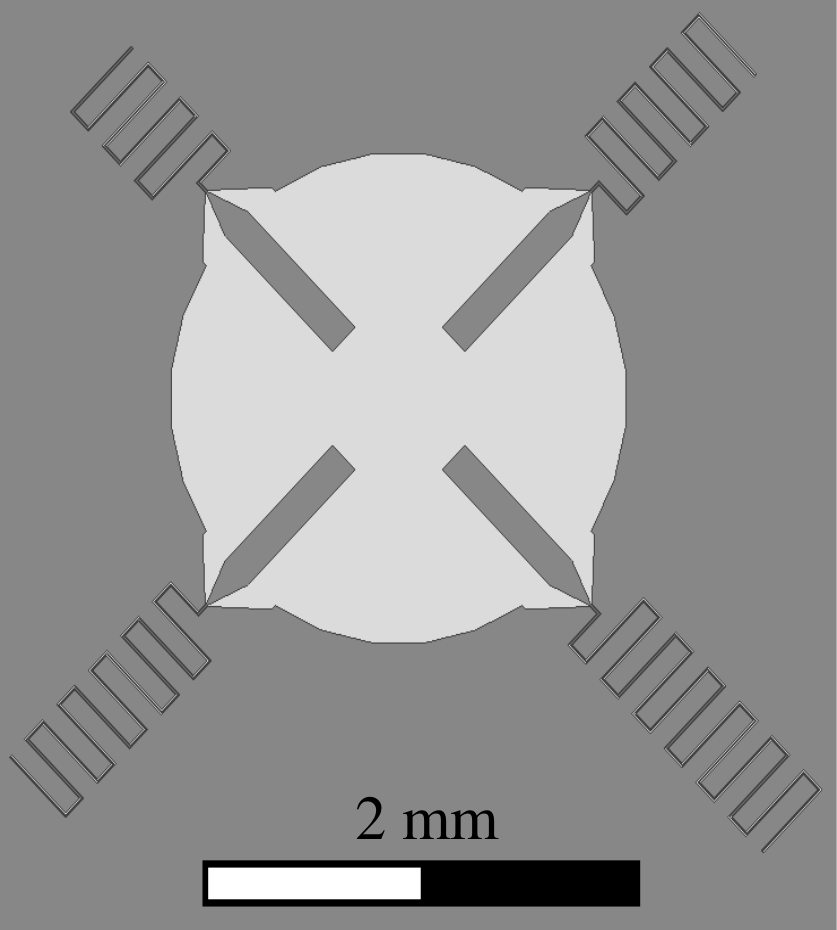}
    \caption{Layout for radio-frequency resonator chip. The resonators are $\lambda/4$ coplanar waveguides connected to a large coupling pin oriented towards the center of the sample. An input-port pin is affixed to the sample box approximately 1 mm above the center of the resonators.}
    \label{figure2}
\end{figure}

Dielectric loss was measured on $a$-Si deposited onto 2 inch diameter wafers at various growth temperatures with thicknesses of either 60 or 180 nm. Two background references were also prepared using the same conditions but with no $a$-Si deposition. After the $a$-Si deposition, an aluminum layer of 100 nm was grown in-situ on each wafer by thermal evaporation at room temperature and 0.2 \AA/s. Four resonators were then patterned with photolithography and an Al wet etch process on each wafer (see Fig.~\ref{figure2}). Measurements of the radio-frequency response were performed in a dilution refrigerator at 10 mK at frequencies ranging from 4 to 7 GHz. Standard cryogenic hygiene was used~\cite{Holland2017}, including 60 dB of attenuation distributed along different temperature stages on the input as well as circulators and HEMT amplifiers on the output. The samples were in a light-tight mu-metal shield to prevent magnetic fields or infrared radiation from interfering with the measurement. In order to increase measurement throughput, rather than designing the samples with a transmission line that requires wirebonding, coupling to the resonators was achieved with a pin that approached within approximately 1 mm of resonator coupling pads. The pin connected to an SMA port and was wired to a T-connector. The resonators were designed to avoid the additional resonance from the T-connector. Once cold, transmission measurements were performed on the resonators as a function of frequency. The curves were normalized and fit to the standard asymmetric resonator transmission equation~\cite{Khalil2012},
\begin{equation}
    S_{21}(\omega) = 1 - \frac{Q_L / \hat{Q}_C}{1 + 2iQ_L \frac{\omega - \omega_0}{\omega_0}}
    \label{equation3}
\end{equation}
where $Q_L$ is the loaded quality factor and its inverse is equal to the sum of the inverted internal quality factor of the resonator, $Q_i$, and the inverted coupling quality factor, $Q_C$, $\omega_0$ is the resonance frequency, and $\hat{Q}_C$ is the complex asymmetric quality factor such that $Q_C^{-1} = Re(\hat{Q}_C^{-1})$.

Dangling bond densities, $\rho_{DB}$, were obtained by electron paramagnetic resonance (EPR) spectroscopy. Measurements were made using a Bruker ELEXSYS E580 EPR spectrometer with an X-band ER 4123D CW-Resonator at 9.36 GHz. Microwave power (1.5 mW) and magnetic field modulation amplitude (5 G) were adjusted for optimum intensity without line shape distortion. Spectra were measured from 3282 to 3383 G. Samples 60, 180 and 300 nm thick were grown on $3 \times 10$ mm$^2$ substrates. A bare substrate was used to determine the background contribution, whereas the samples’ spin densities ($N_S$) were determined by a double integration of the experimental absorption first derivative spectra and comparison to a KCl weak pitch with $N_S = 9.5 \times 10^{12} \pm 5\%$ spins$\,$cm$^{-3}$ and $g = 2.0028 \pm 0.0002$. The estimated relative accuracy of the $N_S$ measurements was $\sim 10\%$. The $a$-Si samples signal was found to be isotropic and have a Landé g-factor of 2.0055, characteristic of dangling bond defects in $a$-Si~\cite{Stutzmann1989}. The spectra show high signal-to-noise ratio, well-resolved lines, and signal saturation for all samples.

\section{Results and discussion}
Two-level systems may interact with an applied field (elastic or electromagnetic) through either relaxation or resonant interactions. When an applied external field of frequency, $\omega$, drives TLSs out of thermal equilibrium, they relax after a time, $\tau$, by exchanging energy with the heat bath via the absorption or emission of thermal phonons with maximal net energy loss when $\omega \tau = 1$. Resonant interactions, on the other hand, occur whenever the energy difference between TLS states match the energy of an applied external field, $E = \hbar \omega$. In general, TLSs excited by an external field will irreversibly emit phonons causing energy loss.

At the frequencies used for the mechanical loss measurements presented in this work, relaxation dominates over resonant interactions above $T = \sqrt[3]{a \omega} \approx 70$ mK, where $a = 10^{-8} \text{s}\,\text{K}^3$, because of the temperature dependence of the TLS relaxation time~\cite{Rau1995}. When the minimum TLS relaxation time $\tau_{min} \omega \ll 1$, TLSs equilibrate within the timescale of an oscillation, and $Q_m^{-1}$ is temperature-independent. This phenomenon is typically observed at temperatures between 0.1 and 10 K, and produces a plateau in $Q_m^{-1}$. The STM predicts that the mechanical loss of a material at this plateau, $Q_{m0}^{-1}$, is given by
\begin{equation}
    Q_{m0}^{-1} = \frac{\pi}{2} \frac{\bar{P} \gamma^2}{\rho v^2}
    \label{equation2}
\end{equation}
where $\bar{P}$ is the TLS density, $\gamma$ is the deformation potential, $\rho$ is the mass density, and $v$ the sound velocity~\cite{Phillips1987}. The deformation potential is defined as $\partial\Delta / 2\partial{u}$, where $\Delta$ is the asymmetry between tunneling states and $u$ is the applied elastic field; in the STM $\gamma$ is also the coupling constant between phonons and TLSs. It is important to note that $\gamma$ is a property of a specific TLS; in deriving Eq.~\ref{equation2}, $\gamma$ is taken as a constant, representing an average over all TLSs.

\begin{figure}
    \centering
    \includegraphics{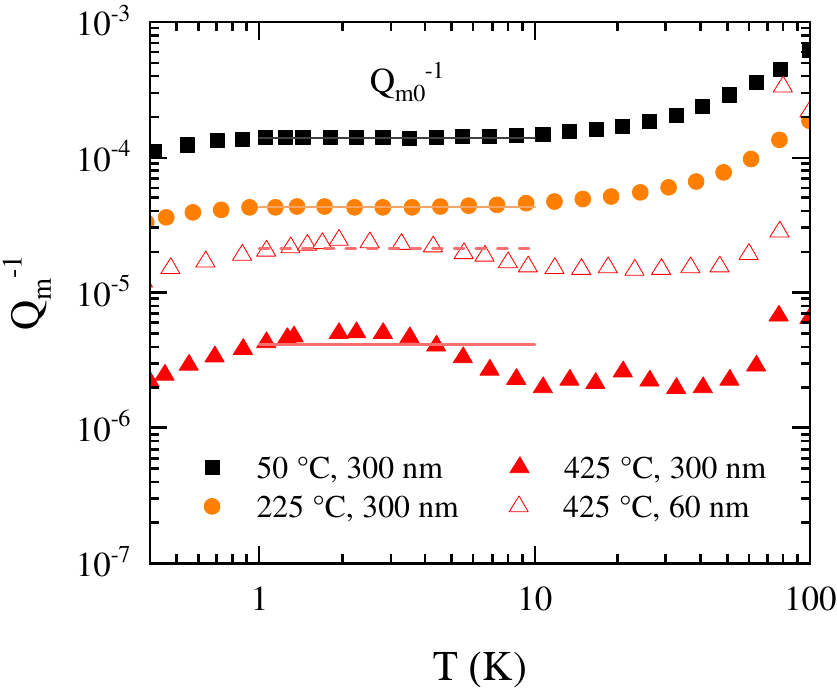}
    \caption{Mechanical loss, $Q_m^{-1}$, as a function of temperature, $T$, for samples grown at 50 (black squares), 225 (orange circles), and 425 \degree C (red triangles). Solid symbols represent 300 nm thick samples and open symbols a 60 nm thick sample. Solid and dashed lines indicate $Q_{m0}^{-1}$ for the 300 and 60 nm samples, respectively.}
    \label{figure1}
\end{figure}

Mechanical loss, $Q_m^{-1}$, for various $a$-Si samples is shown in Fig.~\ref{figure1}. We define the plateau and its value, $Q_{m0}^{-1}$, as the average of the $Q_m^{-1}$ values from 1 to 10 K. Samples with lower losses, such as those grown at 425 \degree C, show in addition a broad peak. This peak is not predicted by the STM and might originate from a non-uniform distribution of barrier heights between neighboring states, which results in an increase of mechanical loss at a particular temperature. In our previous work we suggested that this peak might be caused by contamination effects~\cite{Liu2014}; however, we did not find any signs of contamination in these samples via RBS characterization.

$Q_{m0}^{-1}$ decreases significantly with increasing growth temperature (with other growth parameters held constant) and is reduced by a factor of 34, in agreement with our previous results~\cite{Liu2014}. We also notice, for the two samples grown at 425 \degree C with thicknesses of 60 and 300 nm, that the thinner sample shows nearly 5 times higher $Q_{m0}^{-1}$ than its thicker counterpart. Since $Q_{m0}^{-1}$ is a material intensive property, this indicates that one or more of the STM parameters ($\bar{P}$, $\gamma$, $\rho$, $v$) in $a$-Si films are thickness-dependent. Both $G$ and $\rho$ have been directly measured. The shear modulus, $G$, increases by 67\% with increasing growth temperature (50 to 425\degree C) for 300 nm thick samples, and increases by 5\% with increasing thickness (60 nm to 300 nm) for samples grown at 425 \degree C. The mass density, $\rho$, increases by 11\% with increasing growth temperature (50 to 425\degree C), and changes $\sim$2\% over the thickness range shown in Fig.~\ref{figure1}. As a result, the sound velocity, $v$, increases by 21\% with growth temperature for 300 nm thick samples, and increases by 1\% with thickness for samples grown at 425 \degree C. $\gamma$ has been shown to be proportional to the elastic properties~\cite{Berret1988}, specifically, $\gamma^2 / (\rho v^2)$ is constant, so $\gamma$ is expected to change by 30\%. These results lead to the conclusion that a change in TLS density, $\bar{P}$, with growth parameters is the leading source of the orders of magnitude changes in $Q_{m0}^{-1}$ shown in Fig.~\ref{figure1}.

\begin{figure}
    \centering
    \includegraphics{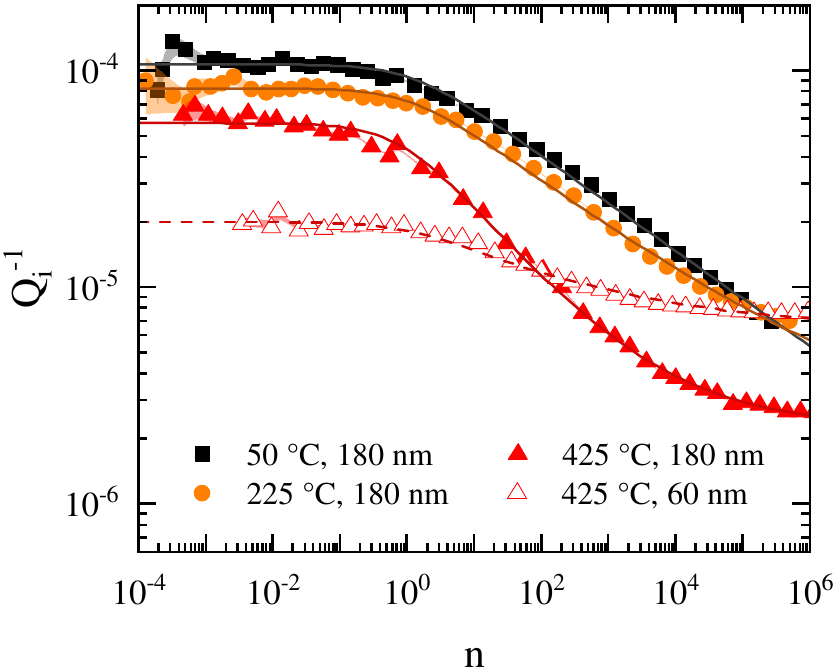}
    \caption{Inverted quality factor, $Q_i^{-1}$, measured at 10 mK as a function of photon number, $n$, (proportional to power) for samples grown at 50 (black squares), 225 (orange circles), and 425 \degree C (red triangles). Solid symbols represent 180 nm thick samples and open symbols a 60 nm thick sample. Solid (180 nm) and dashed (60 nm) lines are fits to the STM power dependence Eq.~\ref{equation4}. Error bars are shown as semi-transparent areas behind the data points.}
    \label{figure3}
\end{figure}

Representative measurements of the inverted quality factor, $Q_i^{-1}$, of the $a$-Si samples, which is proportional to their dielectric loss, is shown in Fig.~\ref{figure3} as a function of the number of photons in the resonator, $n$. We use $n = (2 V^2 Q_L^2)/(50 \hbar \omega_0^2 Q_C)$, where $V$ is the voltage applied over the 50 $\Omega$ terminated transmission line. The loss is fit to the STM power dependence equation:
\begin{equation}
    Q_i^{-1} = \frac{Q_{LP}^{-1}}{(1+\frac{n}{n_0})^{\beta/2}} + Q_{HP}^{-1}
    \label{equation4}
\end{equation}
where $Q_{LP}$ is the low power quality factor, $Q_{HP}$ the high power quality factor, $n_0$ the saturation energy for the resonator’s TLSs, and $\beta$ a fit parameter that in the STM $\beta = 1$~\cite{Phillips1987}, but experimentally is found to be less than 1 (ranging from 0.3 to 0.7, including the present work on $a$-Si)~\cite{Sage2011,Burnett2016}; which has been suggested to indicate interacting TLSs~\cite{Faoro2012,Burnett2014}, or geometrical effects~\cite{Gao2008}.

We performed finite element simulations using ANSYS electronic desktop to determine the stored electric field energy in the $a$-Si layer and multiply $Q_{LP}^{-1}$ by the participation ratio, $PR$, to extract the dielectric loss of the material, $\tan\delta_0 = (Q_{LP} \times PR)^{-1}$ (see Table~\ref{table1})~\cite{Dial2016}. The participation ratio removes thickness dependence, so that $\tan\delta_0$ is independent of the resonator geometry and thus an intensive property. Since the measurements were performed on planar samples, the dielectric loss could be from the $a$-Si layer, the substrate, or the interfaces, limiting the precision of $\tan\delta_0$ of $a$-Si. For each sample we report the error bars that account for this uncertainty by comparing the samples to the bare substrate measurements. In Fig.~\ref{figure3} we note that the power dependence of samples grown at 425 \degree C is much weaker than for samples grown at lower temperatures. Additionally, the 60 nm samples have a larger $Q_{HP}^{-1}$ than the 180 nm samples. The loss for $a$-Si grown at 425 \degree C is likely approaching the weak power dependence limit generally seen in coplanar resonators~\cite{Sage2011,Burnett2016,Khalil2011}.

\begin{table*}[]
    \setlength{\tabcolsep}{8pt}
    \centering
    \caption{Summary of data: growth temperature $\text{T}_\text{S}$, film thickness $t$, mass density $\rho$, dangling bond density $\rho_{DB}$, shear modulus $G$, sound velocity $v$, mechanical loss $Q_{m0}^{-1}$ at 1 to 10 K and 5 kHz, and loss tangent $\tan\delta_0$ at 10 mK and 5 GHz.}
    \begin{tabular}{c c c c c c c c} \hline \hline
    $\text{T}_\text{S}$ & $t$ & $\rho$ & $\rho_{DB}$ & $G$ & $v$ & $Q_{m0}^{-1}$ & $\tan\delta_0$ \\
     &  &  & $\times 10^{18}$ &  & $\times 10^3$ & $\times 10^{-6}$ & $\times 10^{-4}$ \\
    (\degree C) & (nm) & (g$\,$cm$^{-3}$) & (spins$\,$g$^{-1}$) & (GPa) & (m$\,$s$^{-1}$) &  &  \\ \hline
    50 & 181.6$\pm$2.3 & 2.06$\pm$0.04 & 2.03$\pm$0.19 & - & - & - & 19.8$\pm$1.3 \\
    50 & 317.4$\pm$1.6 & 2.08$\pm$0.04 & 2.04$\pm$0.19 & 36.8$\pm$0.2 & 4.2$\pm$0.1 & 140.0$\pm$2.8 & - \\ 
    225 & 173.9$\pm$2.1 & 2.19$\pm$0.04 & 1.67$\pm$0.16 & - & - & - & 13.9$\pm$1.3 \\
    225 & 310.0$\pm$1.8 & 2.20$\pm$0.04 & 1.67$\pm$0.16 & 48.7$\pm$0.2 & 4.7$\pm$0.1 & 43.1$\pm$0.9 & - \\ 
    425 & 169.9$\pm$1.9 & 2.29$\pm$0.04 & 1.16$\pm$0.11 & - & - & - & 8.6$\pm$1.3 \\
    425 & 299.2$\pm$1.4 & 2.30$\pm$0.04 & 1.20$\pm$0.12 & 61.3$\pm$0.2 & 5.2$\pm$0.1 & 4.1$\pm$0.1 & - \\ 
    425 & 59.7$\pm$3.1 & 2.28$\pm$0.05 & 1.05$\pm$0.11 & - & - & - & 3.3$\pm$3.5 \\
    425 & 59.2$\pm$1.2 & 2.27$\pm$0.05 & 1.05$\pm$0.11 & 58.4$\pm$0.6 & 5.1$\pm$0.1 & 21.3$\pm$0.4 & - \\ \hline \hline
    \end{tabular}
    \label{table1}
\end{table*}

The values of $\tan\delta_0$ shown in Table~\ref{table1} indicate that the dielectric loss is strongly thickness-dependent in $a$-Si films. Thus, mechanical and dielectric loss results are both thickness-dependent, which suggests that the structures responsible for TLSs change with thickness. However, the trends are opposite: mechanical loss is lower for thicker films, whereas dielectric loss is lower for thinner films.

The STM predicts that the TLS density, $\bar{P}$, is proportional to the dielectric loss, $\tan\delta_0$, when coupled to dipole moments via electric fields~\cite{Phillips1987}. Dielectric loss is frequently measured using superconducting resonators or qubits at GHz frequencies, and at mK temperatures to avoid TLS saturation~\cite{Sarabi2016,Brehm2017,Niepce2020}; at $\hbar \omega / (2 k_B T) \simeq 12 \gg 1$, resonant interactions are the dominant loss mechanism. In this regime, the dielectric loss, $\tan\delta_0$, is defined by
\begin{equation}
    \tan\delta_0 = \pi \frac{\bar{P} p_0^2}{3\epsilon}
    \label{equation5}
\end{equation}
where $p_0$ is the (average) TLS dipole moment, and the permittivity $\epsilon = \epsilon_0 \epsilon_r$. We used $\epsilon_0 = 8.85 \times 10^{-12}$  F$\,$m$^{-1}$ for the electric constant, and $\epsilon_r = 11.45$ for the silicon relative permittivity, obtained at 10 K in the GHz range~\cite{Krupka2006}. As with $\gamma$, it is important to note that the TLS dipole moment, $p_0$, is a property of a specific TLS; in deriving the above, $p_0$ is also taken as a constant, representing an average over all TLSs.

To understand the possible mechanisms giving rise to the mechanical and dielectric loss results presented in this work, we consider the samples’ density and dangling bond density, which provide information about the number of atoms and unpaired electrons per unit volume present in each sample. 

The density, $\rho$, shown in Table~\ref{table1}, increases with growth temperature and with film thickness, primarily caused by a reduction of the density of open-volume defects, i.e., nanovoids, as found in our previous work on $a$-Si~\cite{Jacks2018,Jacks2020}. The difference in density between 180 and 300 nm is however negligible at all growth temperatures. Dangling bond density, $\rho_{DB}$, decreases with growth temperature, an indication that denser films contain fewer dangling bond defects, but shows little dependence on thickness, except for films grown at the higher temperature, where $\rho_{DB}$ increases slightly with increasing film thickness~\cite{Jacks2018,Jacks2020}. These results make clear that the relationship between atomic density and dangling bond density is not straightforward. 

Mechanical loss results presented in this work are dominated by relaxation interactions, whereas dielectric loss results are dominated by resonant interactions, which hinders a direct comparison between $Q_{m0}^{-1}$ and $\tan\delta_0$. However, in the framework of the STM, the TLS distribution function $f(\Delta,\lambda)$ is only a function of the asymmetry between states, $\Delta$, and the tunneling parameter, $\lambda$. The main assumptions of the STM are that $\Delta$ and $\lambda$ are independent of each other and uniformly distributed, therefore $f(\Delta,\lambda) \approx \bar{P}$, where $\bar{P}$ is the TLS density, an energy-independent parameter. Comparisons of $\bar{P}$ obtained from mechanical or dielectric loss results can therefore be done under the assumptions of the STM, despite their different regimes.

\begin{figure}
    \centering
    \includegraphics{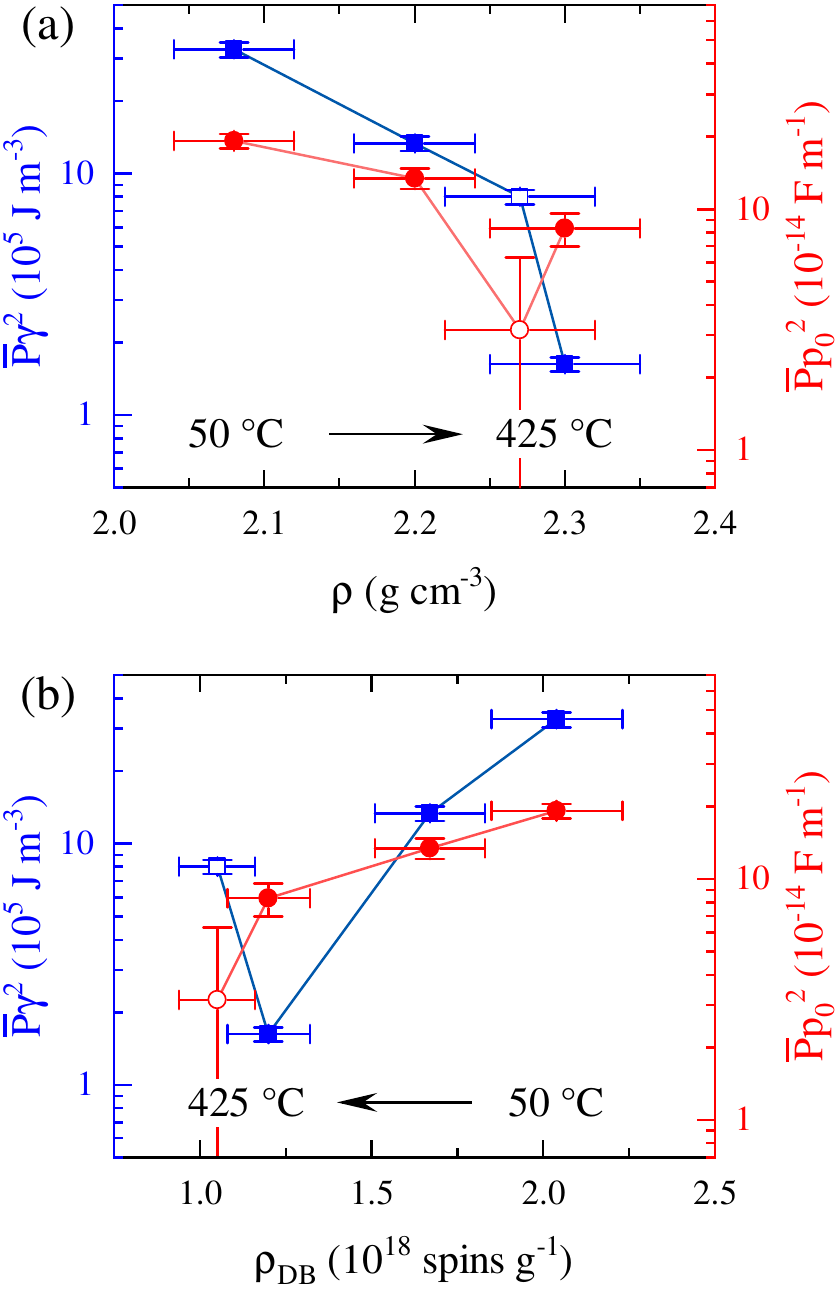}
    \caption{Mechanical loss-derived quantity, $\bar{P}\gamma^2$, (blue left axis and squares) and dielectric loss-derived quantity, $\bar{P}p_0^2$, (red right axis and circles) as a function of (a) density, $\rho$, and (b) dangling bonds density, $\rho_{DB}$. Both quantities are respectively derived from Eqs.~\ref{equation2} and \ref{equation5} using data reported in Table~\ref{table1}. Closed symbols are thick films (180 or 300 nm) and open symbols are thin films (60 nm). Black arrows indicate the growth temperature increase for the samples reported in these plots. Crystalline Si density is 2.33 g/cm$^3$; the films grown at higher temperature approach this density, despite remaining fully amorphous.}
    \label{figure4}
\end{figure}

The coupling constant, $\gamma$, and dipole moment, $p_0$, are neither well understood nor measured for many materials, and are additionally TLS-specific, 
which makes it more difficult to decouple them from $\bar{P}$ (see Eqs.~\ref{equation2} and \ref{equation5}). For this reason, Fig.~\hyperref[figure4]{\ref*{figure4}(a)} shows the parameters $\bar{P}\gamma^2$ (mechanical loss) and $\bar{P}p_0^2$ (dielectric loss) as a function of density, $\rho$, and Fig.~\hyperref[figure4]{\ref*{figure4}(b)} shows $\bar{P}\gamma^2$ and $\bar{P}p_0^2$ as a function of dangling bond density, $\rho_{DB}$. Considering first the effect of growth temperature only (solid symbols), both $\bar{P}\gamma^2$ and $\bar{P}p_0^2$ decrease as growth temperature increases. However, $\bar{P}\gamma^2$ and $\bar{P}p_0^2$ are not proportional to each other, and when the effect of thickness is included (open symbols), the lack of proportionality becomes even more striking. Fig.~\hyperref[figure4]{\ref*{figure4}(a)} shows that $\bar{P}\gamma^2$ has a systematic correlation with density (whether caused by thickness or growth temperature) while $\bar{P}p_0^2$ does not; the 60 nm sample grown at 425 \degree C shows the lowest dielectric loss value despite reporting an intermediate density. Fig.~\hyperref[figure4]{\ref*{figure4}(b)}, by contrast, shows that $\bar{P}p_0^2$ is monotonic with the density of dangling bonds, while $\bar{P}\gamma^2$ is not; the 300 nm sample grown at 425 \degree C shows the lowest mechanical loss value despite reporting an intermediate dangling bond density. We note that an increase in density and a reduction in dangling bonds density each represent a reduction of some type of structural defects. Our results show that mechanical loss, $Q_{m0}^{-1}$, correlates with density (an increase in $\rho$ by 11\% is associated with a factor of 34 reduction in $Q_{m0}^{-1}$), while dielectric loss, $\tan\delta_0$, correlates with dangling bond density (a reduction in $\rho_{DB}$ by a factor of 2 is associated with a factor of 6 reduction in $\tan\delta_0$).

Comparing only the thick samples, the factor of 34 reduction of mechanical loss with increased growth temperature is in stark contrast to the factor of 2.3 reduction of dielectric loss. This difference can be explained by two different hypotheses within the framework of the STM: (i) the density of TLS, $\bar{P}$, is a single quantity for a given sample, and differences between mechanical and dielectric loss are due to changes in coupling constant, $\gamma$, and dipole moment, $p_0$, or (ii) there are different species of TLSs yielding independent values of $\bar{P}$ for mechanical and dielectric loss.

In the framework of the first hypothesis, (i), the same TLS density, $\bar{P}$, contributes to both types of losses. The large difference (34 vs 2.3) between the dependency of $Q_{m0}^{-1}$ and $\tan\delta_0$ on growth temperature (see Table~\ref{table1}) is then explained by changes of $\gamma$ and $p_0$ with growth temperature. The relationship between coupling constant and elastic properties,  $\gamma^2 / (\rho v^2)$, is a constant for a wide variety of glasses~\cite{Berret1988}, which implies that $\bar{P}$ is reduced by a factor of 34 with increasing growth temperature from 50 to 425 \degree C (see Eq.~\ref{equation2}). If $\bar{P}$ decreases by a factor of 34, then $p_0^2 / \epsilon$ must increase 15 times (Eq.~\ref{equation5}) to account for the smaller 2.3 times decrease in $\tan\delta_0$. If the permittivity, $\epsilon$, were changing, we would observe changes in the resonant frequency of the resonators. To reconcile these results with the predictions made by the STM, the average TLS dipole moment, $p_0$, would therefore need to increase almost 4 times with increasing growth temperature, to allow for a TLS density reduction of 34 times, an unlikely scenario since neither the charge nor separation of atoms participating in the tunneling are likely to increase fourfold. Hypothesis (i) therefore seems quite unlikely.

The second hypothesis, (ii), concerns the nature of the structural defects that give rise to TLSs. Two different subsets of defects, one that strongly couples only to elastic fields, and another that strongly couples to electric fields, could yield two different densities of TLSs: $\bar{P}_m$ and $\bar{P}_e$, respectively. In this case, the coupling constant and dipole moment could remain almost independent of growth parameters. Assuming a weak dependence of $\gamma^2 / (\rho v^2)$ and $p_0^2 / \epsilon$ on growth conditions, the data suggest a reduction of $\bar{P}_e$ of only 2.3 times with increasing growth temperature, about 15 times smaller than the reduction in $\bar{P}_m$.

Since mechanical losses correlate with atomic density, while dielectric losses correlate with dangling bond density, considering different $\bar{P}_m$ and $\bar{P}_e$ is a natural concept. We have previously suggested that TLSs that yield $\bar{P}_m$ are spatially correlated with nanovoids or low density regions~\cite{Liu2014}, and have shown that nanovoids are the structural defect associated with the observed reduced atomic density in $a$-Si films~\cite{Jacks2018,Jacks2020}. By analogy, we suggest that the TLSs that yield $\bar{P}_e$ are connected in some way with dangling bonds; while it is unlikely that dangling bonds are directly producing TLSs, there may be a correlation spatially.

Previous work (see Table I in Ref.~\citep{Pohl2002}) supports this hypothesis of different types of TLSs. Different amounts of OH impurities in silica had no effect in elastic measurements~\cite{Hunklinger1975}, whereas dielectric measurements were clearly dependent on the OH concentration~\cite{Schickfus1976}. It was concluded that not all elastically coupled TLSs interact electrically~\cite{Arnold1976}. Our results show that even in a material without impurities, elastically interacting TLSs can be considerably reduced without large reductions in the dielectric TLS density, $\bar{P}_e$. Several models have been proposed to understand the apparent mismatch between thermal, acoustic, and electric properties in the context of the predictions of the STM. These approaches include ``anomalous'' TLSs that modify the distribution of relaxation times~\cite{Black1978}, two different types of TLSs that couple weakly or strongly with elastic or electric fields~\cite{Schechter2013,Sarabi2016}, and self-interacting TLSs that modify the coupling between phonons and TLSs~\cite{Carruzzo2020}.

The mechanical and dielectrics loss results presented in this paper, obtained for two quite different frequency ranges (kHz and GHz), could also support the idea that the TLS density, $\bar{P}$, is not energy independent nor uniformly distributed. A frequency dependence of the distribution of states would significantly modify the STM; such a model, which predicts dephasing and noise in superconducting microresonators due to TLSs in amorphous materials, has recently been explored~\cite{Faoro2015}.

\section{Conclusions}
Amorphous silicon films show orders of magnitude reduction of the mechanical loss plateau, $Q_{m0}^{-1}$, with increased growth temperature at fixed thickness, whereas dielectric loss, $\tan\delta_0$, is only reduced by a factor of 2.3. Furthermore, mechanical loss is correlated with atomic density, while dielectric loss is correlated with dangling bond density.  The most plausible explanation for these data is that there are different types of TLSs that interact with external fields (elastic and electromagnetic) through different mechanisms.

A better understanding of the TLS-phonon coupling constant and TLS dipole moment, their relationship with atomic structure, and how the defects that give rise to TLSs interact through different dissipation mechanisms are needed to disentangle the underlying physics of the anomalous properties of disordered solids at low temperatures.

\begin{acknowledgments}
The UCB part of this work was supported by NSF DMR-0907724, 1508828, and 1809498, and by the Joint Center for Artificial Photosynthesis, a DOE Energy Innovation Hub, supported through the Office of Science of U.S. Department of Energy under Award Number DE-SC0004993. MMR thanks R. Chatterjee and J. K. Cooper for assistance with EPR measurements. The LLNL portion of this work was performed under the auspices of the U.S. Department of Energy by Lawrence Livermore National Laboratory under Contract DE-AC52-07NA27344 and was supported by the Laboratory Directed Research and Development under Grant No. 16-SI-004 and 20-ERD-010. This work was also partially supported by the Office of Naval Research at NRL.
\end{acknowledgments}

\bibliography{references}

\end{document}